\documentclass[a4paper]{jpconf}

\usepackage[ansinew]{inputenc}
\usepackage[T1]{fontenc}
\usepackage[english]{babel}
\usepackage[usenames,dvipsnames]{color}
\usepackage{graphicx}
\usepackage{amsmath}
\usepackage{amssymb}
\usepackage{hyperref}
\usepackage[numbers,sort&compress,merge]{natbib}
\usepackage{float}
\usepackage{multirow}

\newcommand{\code}[1]{#1}
\newcommand{\cxx}{\code{C++}}

\newcommand{\inclxx}{\code{INCL++}}

\newcommand{\incl}{\code{INCL}}

\newcommand{\isabel}{\code{Isabel}}
\newcommand{\bertini}{\code{Bertini}}

\newcommand{\abla}{\code{ABLA07}}

\newcommand{\proton}{\textit{p}}

\newcommand{\lead}{$^\text{208}$Pb}
\newcommand{\calcium}{$^\text{40}$Ca}

\begin{document}
\title{Shell structure and few-nucleon removal in intranuclear cascade}

\author{D~Mancusi$^{1}$, A~Boudard$^{1}$, J~Carbonell$^{1}$, J~Cugnon$^{2}$,
  J-C~David$^{1}$ and S~Leray$^{1}$}

\address{$^{1}$ CEA, Centre de Saclay, IRFU/Service de Physique Nucl\'eaire,
  F-91191 Gif-sur-Yvette, France}
\address{$^{2}$ University of Li\`ege, AGO Department, all\'ee du 6 Ao\^ut 17,
  b\^at. B5, B-4000 Li\`ege 1, Belgium}

\ead{davide.mancusi@cea.fr}

\begin{abstract}
  It is well known that intranuclear-cascade models generally overestimate the
  cross sections for one-proton removal from heavy, stable nuclei by a
  high-energy proton beam, but they yield reasonable predictions for one-neutron
  removal from the same nuclei and for one-nucleon removal from light
  targets. We use simple shell-model calculations to investigate the reasons of
  this deficiency. We find that a correct description of the neutron skin and of
  the energy density in the nuclear surface is crucial for the aforementioned
  observables. Neither ingredient is sufficient if taken separately.
\end{abstract}

\section{Introduction}
\label{sec:introduction}

Nuclear reactions between high-energy ($\gtrsim150$~MeV) nucleons or hadrons and
nuclei are usually described by means of \emph{intranuclear-cascade} (INC)
models \cite{serber-reactions}. In this framework, the projectile is assumed to
initiate an avalanche of binary collisions with the nucleons of the target,
which can lead to the emission of energetic particles. The nature of INC models
is essentially classical. It is typically assumed that nucleons are perfectly
localised in phase space and are bound by an average, constant potential;
moreover, it is assumed that subsequent elementary collisions are independent.

It was realized some time ago that INC models systematically fail to describe
inclusive cross sections for the removals of few nucleons \cite[see
e.g.][]{jacob-p2p,audirac-evaporation_cost}. This is especially surprising in
view of the fact that these observables are associated with peripheral reactions
and mostly involve collisions between quasi-free nucleons; one would therefore
expect intranuclear cascade to provide an accurate description of this
particular dynamics. This puzzling result has been known for many years now, but
no convincing explanation has ever been put forward.

We will show that the few-nucleon removal process at high energy is sensitive to
the the description of the nuclear surface, which we draw from a simple
shell-model calculation. We will show that the predictions of an INC model
\cite{boudard-incl4.6} can be substantially improved by casting the shell-model
calculation results in a form adaptable to the nuclear model underlying INC.

\section{Model description}\label{sec:model-description}

It is generally assumed that the first stage of high-energy nucleon-nucleus
reactions can be described as an avalanche of independent binary
collisions. 
The nuclear model underlying INC is essentially classical, with the addition of
a few suitable ingredients that mimic intrinsically quantum-mechanical features
of the initial condition and of the dynamics. 
At the end of the intranuclear cascade, an excited remnant is left. The
de-excitation of this nucleus is typically described by a statistical
de-excitation model. 

In what follows, we shall make explicit reference to the Li\`ege Intranuclear
Cascade model \cite[\incl,][]{boudard-incl4.6} and the \abla\ statistical
de-excitation model \cite{kelic-abla07}. The \incl/\abla\ coupling is in general
quite successful at describing a vast number of observables in nucleon-induced
reactions at incident energies between $\sim60$ and $3000$~MeV
\cite{leray-intercomparison,*intercomparison-website}. The work described
hereafter was performed with the latest \cxx\ version of \incl\
\cite[\inclxx,][]{mancusi-inclxx}. 

The \incl\ model is peculiar in that it explicitly tracks the motion of all the
nucleons in the system, which are assumed to move freely in a square potential
well. The radius of the well is not the same for all nucleons, but it is rather
a function $R(p)$ of the absolute value of the particle momentum (which is a
conserved quantity in absence of collisions). The initial particle momenta are
uniformly distributed inside a sphere of radius $p_F=270$~MeV${}/c$.
The relation between momentum and radius of the potential well is such that the
space density distribution is given by a suitable Saxon-Woods parametrisation;
moreover, the nuclear surface is predominantly populated by nucleons whose
energy is close to the Fermi energy \cite{boudard-incl}.


\section{One-nucleon-removal cross sections}
\label{sec:one-nucleon-removal}

\begin{figure}
  \centering
  \begin{minipage}{0.49\linewidth}
    \includegraphics[width=\linewidth]{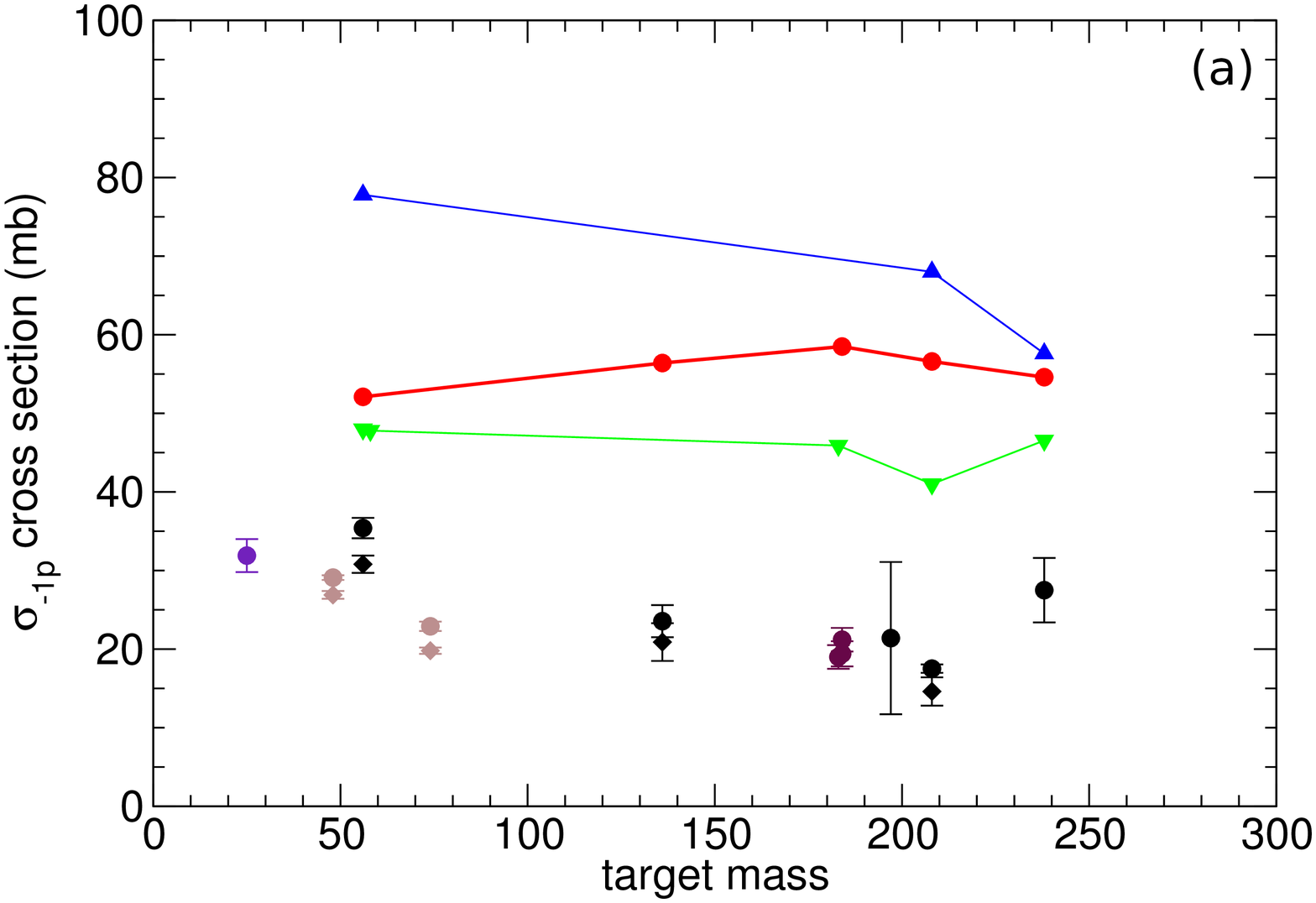}
  \end{minipage}
  \begin{minipage}{0.49\linewidth}
    \includegraphics[width=\linewidth]{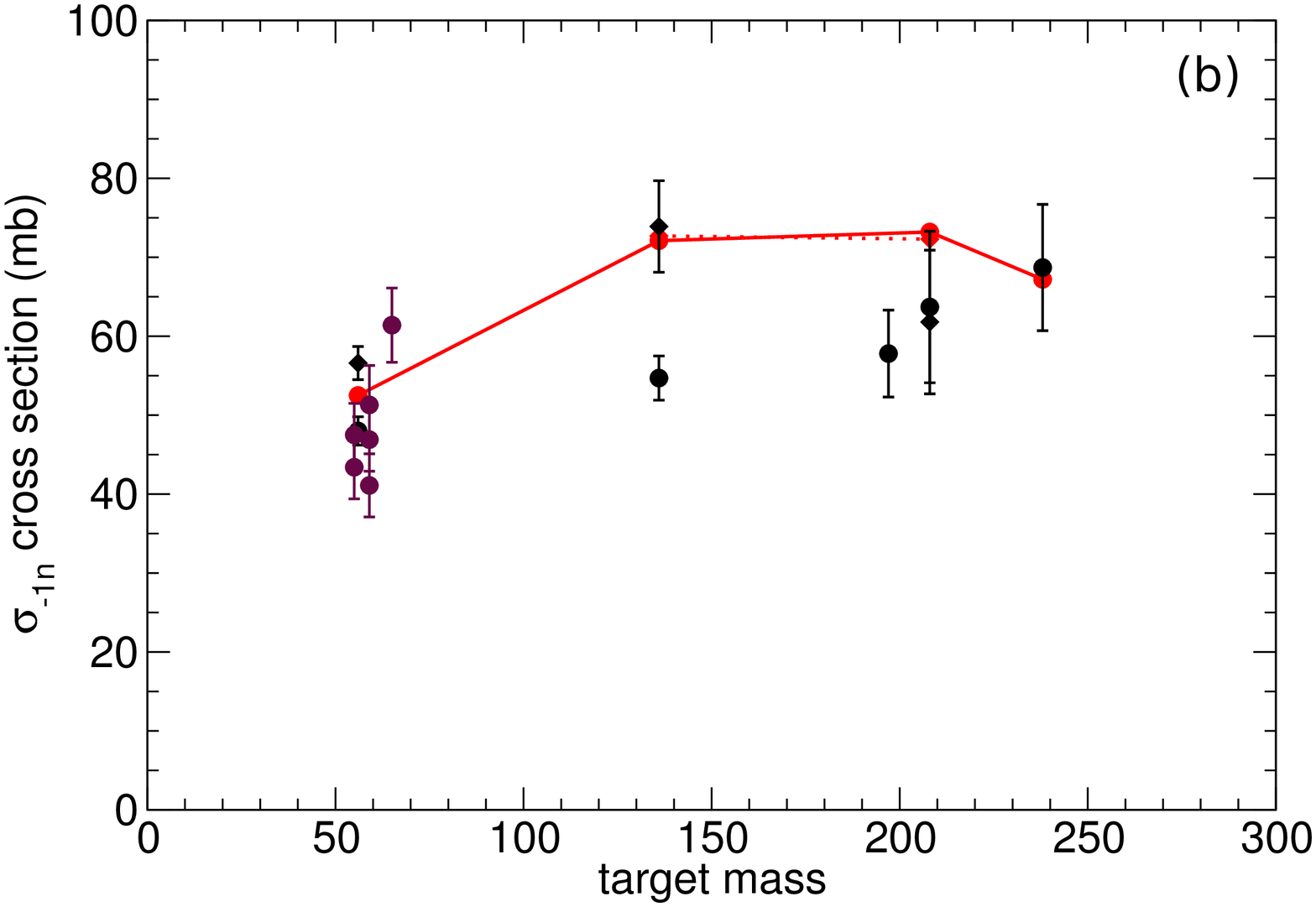}
  \end{minipage}
  \caption{Experimental data for one-proton- (a) and one-neutron-removal cross
    sections (b) in proton-nucleus reactions above 500~MeV incident energy, as a
    function of the target mass. Diamonds refer to experimental beam energies
    around 500~MeV, while circles represent energies between 800 and
    1000~MeV. The solid curves represent calculations with \incl\ (red),
    \isabel\ (blue) \cite{yariv-isabel1} and \bertini\
    \cite{bertini-cascade1,*bertini-cascade2} (green) at 1000~MeV. Experimental
    data taken from
    Refs.~\citenum{villagrasa-fe,rejmund-gold,giot-xe_500MeV,napolitani-xe_xsec,audouin-lead,enqvist-lead,taieb-u,titarenko-indc,michel-nuclide,jacob-p2p,reeder-mg25_p2p}.\label{fig:world-data}}
\end{figure}

Figure~\ref{fig:world-data} shows the experimental data for one-nucleon removal
in proton-induced reactions at energies $\gtrsim500$~MeV, as a function of the
target mass (all targets are $\beta$-stable). Calculations with \incl/\abla\ are
shown for comparison. It is clear that the model predictions are in the right
ballpark for neutron removal, but they overestimate the proton-removal data by a
factor that can be as large as 3--4 for heavy
nuclei. 
Note also that other cascade models similarly overestimate the proton-removal
cross sections.
Figure~\ref{fig:world-data} suggests that INC models might
be affected by a fundamental defect. It is however rather surprising that the
deficiency clearly manifests itself in proton removal, but neutron removal seems
unaffected. 

The analysis of the model calculations indicates that one-proton removal is
dominated (about 90\% of the cross section) by events with only one
proton-proton collision. The two protons leave the nucleus, which however
retains some excitation energy. If only one collision took place, the excitation
energy is given by the energy of the proton hole, i.e.\ the difference between
the Fermi energy and the energy of the proton that was ejected. This excitation
energy is evacuated during the de-excitation stage by neutron evaporation.
If the excitation energy is lower than the neutron separation
energy, no particle will be evaporated and the energy will be evacuated as gamma
rays; in this case the final (observed) residue will therefore be the target
nucleus minus one proton. If the excitation energy allows for neutron
evaporation, the final residue will be lighter (target minus one proton minus $x$
neutrons).
The one-proton-removal cross sections are therefore extremely sensitive to the
excitation energy left in the nucleus after the
cascade. 
Note that there is a subtle difference between one-proton and one-neutron
removal.  One-neutron removal can be realized in two ways: either as a neutron
ejection during INC followed by no evaporation (this is analogous to the
proton-removal mechanism), or as no neutron ejection during INC followed by
evaporation of one
neutron. 

Our results are essentially independent of the choice of the de-excitation
model, since all of them employ very similar separation energies for stable
nuclei. 
Comparison with the experimental data (Fig.~\ref{fig:world-data}) seems to
suggest that INC underestimates the excitation energy associated with the
ejection of a proton; larger excitation energies would lead to increased neutron
evaporation and would therefore reduce the one-proton-removal cross section.



\section{Refinement of the INC nuclear model}
\label{sec:refined-inc-model}

We mentioned at the end of Section~\ref{sec:model-description} that the nuclear
surface is predominantly populated by nucleons whose energy is close to the
Fermi energy. The ejection of one such nucleon during INC results in little
excitation energy for the cascade
remnant. 
However, even deeply-bound nucleons have a non-vanishing probability to be found
in the nuclear surface; this aspect is usually neglected by INC models. 
%
Another detail that is usually neglected in the INC picture is the
presence of neutron (or proton) skins in certain nuclei, such as \lead. 
For surface reactions, this means that the local neutron density is several
times larger than the proton density, leading to an enhanced probability for
proton-neutron collisions.

\subsection{Shell-model calculations}
\label{sec:shell-model-calc}

We have estimated the magnitude of both the effects above with a simple
shell-model calculation. We assumed a central Saxon-Woods nuclear potential with
a spin-orbit term and a Coulomb term for the protons \cite{blomqvist-shell}. We
numerically solved the radial part of the Schr\"{o}dinger equation and
determined the eigenfunctions and the eigenvalues of the bound states.  The
single-particle energies correctly reproduce the energies of the particle-hole
states in $^{207,209}$Pb and $^{207}$Tl,Bi.



We would like to use the shell-model proton and neutron densities as inputs for
our INC calculation; however, the particle densities in \incl\ cannot be given
by an arbitrary function, so we must somehow adapt the shell-model densities. We
chose to fit them with Saxon-Woods distributions (shown in
Fig.~\ref{fig:density} as dashed lines). The best-fit parameters show that the
shell-model densities exhibit a neutron skin in
\lead. 
We have thus decoupled the \incl\ parameters describing the neutron space
density from those describing the proton space density. The proton densities
have not been modified (because they are already given by fits to the
experimental charge radii), but the neutron parameters have been adjusted by the
skin thicknesses resulting from the fit shown in Fig.~\ref{fig:density}.

\begin{table}
\begin{minipage}{\linewidth}
  \begin{minipage}{0.48\linewidth}
    \begin{figure}[H]
      \includegraphics[width=\linewidth]{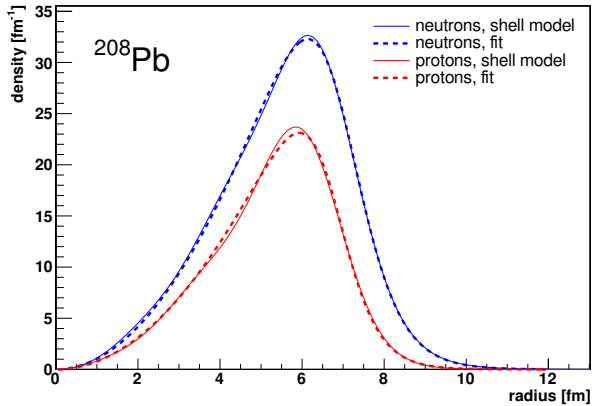}
      \caption{Proton (red) and neutron (blue) densities for \lead. The thin
        solid lines represent the result of the shell-model calculation, while
        the thick dashed lines are Saxon-Woods fits.\label{fig:density}}
    \end{figure}
  \end{minipage}\hfill%
  \begin{minipage}{0.48\linewidth}
    \begin{table}[H]
      \caption{Cross sections for
        one-nucleon removal in 1-GeV \proton-nucleus reactions, with the
        following model variants: (a) standard, (b) standard plus neutron skin,
        (c) standard plus surface fuzziness, (d) standard plus neutron skin and
        surface fuzziness. Experimental data are taken from
        Refs.~\citenum{enqvist-lead,chen-ca40}.\label{tab:cross-sections}}
      \centering
      \setcounter{footnote}{1}
      \lineup
      \begin{tabular}{ccccc}
        \br
        & \multicolumn{2}{c}{\calcium} &  \multicolumn{2}{c}{\lead}\\
        & $-1p$ & $-1n$ &  $-1p$ & $-1n$ \\
        \mr
        (a) & $59.8$ & $46.4$ & $59.5$ & $\082.1$\\
        (b) & $58.8$ & $41.4$ & $50.9$ & $112.0$\\
        (c) & $51.6$ & $38.3$ & $42.1$ & $\063.4$\\
        (d) & $51.9$ & $35.3$ & $33.6$ & $\083.8$\\
        \mr
        \multirow{2}{*}{exp\footnote{The experimental values for \calcium\ refer to
            an incident energy of 763~MeV.}} & $\m54.7$ & $\m29.8$ & $\m17.6$ & $\m63.7$\\
        & $\pm\07.9$ & $\pm\06.4$ & $\pm\00.5$ & $\pm\09.6$\\
        \br
      \end{tabular}
    \end{table}
  \end{minipage}
\end{minipage}
\end{table}

We have explained in the previous section that the outcome of single-collision
cascades is sensitive to the energy of the ejected nucleon. We assume that the
probability that a collision ejects a nucleon from a given shell is proportional
to the local density of the shell orbital. Furthermore, we neglect rearrangement
of the other nucleons in the Fermi sea after the collision; this amounts to
assuming that the excitation energy of the hole is simply given by the depth of
the hole, measured from the Fermi energy.



With these assumptions, we can estimate the mean and RMS excitation energies
that are left in the nucleus if a hole is punched in the Fermi sea at a certain
distance from the center. These quantities are plotted in
Fig.~\ref{fig:estar-mean-rms} for the shell-model calculation and for the
standard \incl\ nuclear model ($f=0$).



It is clear from the results displayed in this picture that the standard \incl\
nuclear model yields mean and RMS excitation energies that are quite different
from those resulting from the shell model. In the surface region, the proton
mean and RMS values from \incl\ are sensibly lower than their shell-model
counterparts, which seems to confirm that the excitation energy associated with
the ejection of a proton is underestimated by \incl.

\subsection{Surface fuzziness}
\label{sec:surface-fuzziness}

We mentioned in Section~\ref{sec:model-description} that an \incl\ nucleon moves
in a square-well potential whose radius $R(p)$ depends on the nucleon
momentum. The function $R(p)$ is uniquely determined by the choice of the space
density $\rho(r)$ and by the assumption that nucleon momenta are uniformly
distributed in a sharp-surface Fermi sphere. We have shown above that this
construction results in excitation energies for one-collision reactions that are
much smaller than those resulting from the shell model and, arguably, than those
suggested by the available experimental data.

We refine the \incl\ nuclear model by making $R(p)$ into a random variable. We
introduce a \emph{fuzziness parameter} $f$ ($0\leq f\leq1$) and a \emph{fuzzy}
square-well radius $R(p,f)$. The precise definition of $R(p,f)$ is outside the
scope of this short paper, but suffice it to say the following: first, for $f=0$
we recover the standard sharp correlation ($R(p,0)=R(p)$). Second, for given
values of $p$ and $f$, $R(p,f)$ is a random variable that describes the radius
of the square well. The fluctuations in $R(p,f)$ are small if $f$ is close to
zero and they are large if $f$ is close to one. Moreover, the fluctuations are
constructed in such a way that the space density is still given by $\rho(r)$ and
the momentum density is still given by a sharp-surface Fermi sphere.

The construction of the refined \incl\ nucleus is analogous to the standard
preparation algorithm \cite{boudard-incl}. The only difference is that the radius
of the square-well potential is no longer in one-to-one correspondence with the
nucleon momentum.

\begin{figure}
  \centering
  \begin{minipage}{0.66\linewidth}
    \includegraphics[width=\linewidth]{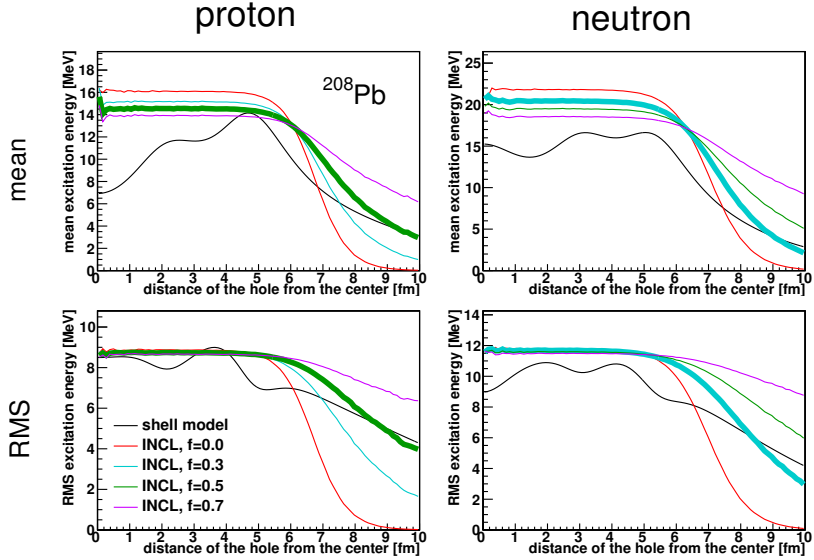}
  \end{minipage}\hfill%
  \begin{minipage}{0.32\linewidth}
    \caption{Mean (top) and root-mean-square (bottom) excitation energy induced
      by the presence of a proton (left) or neutron hole (right) in \lead, as
      functions of the hole position. The dashed lines represent the shell-model
      result, while the solid lines are the results generated by the \incl\
      nuclear model for different values of the fuzziness parameter $f$. The
      thick lines represent the selected parameter
      values.\label{fig:estar-mean-rms}}
  \end{minipage}
\end{figure}

The refined nuclear model introduces fluctuations in the space distribution of
nucleons with a given energy; equivalently, it introduces additional energy
fluctuations for the nucleons found at a given
position. Figure~\ref{fig:estar-mean-rms} indeed demonstrates that the average
and RMS excitation energies for surface holes \emph{increase} for increasing
surface fuzziness, i.e.\ for increasing fluctuations. No value of the fuzziness
parameter yields a good fit to the shell-model result, even if one limits
oneself to the surface region. There is some degree of subjectivity in the
choice of the best-fit values, which are taken to be $f=0.5$ for protons and
$f=0.3$ for neutrons. For \calcium\ (not shown), the best-fit value was taken to
be $f=0.3$ for both protons and neutrons.

Summarizing, we have refined the \incl\ nuclear model in two respects. First, we
have introduced a neutron skin, as described in
Section~\ref{sec:shell-model-calc}. Second, we have introduced surface
fuzziness, which increases the energy content of the nuclear surface and the
probability for deep-nucleon removal in surface collisions. In the framework of
the shell model, this effect is genuinely quantum-mechanical and is due to the
penetration of the wavefunction in the classically forbidden region.

\section{Results and conclusions}\label{sec:results-conclusions}

We turn now to the analysis of the results of the refined INC
model. Table~\ref{tab:cross-sections} shows how the neutron skin
and the surface fuzziness affect the one-nucleon-removal cross sections in 1-GeV
\proton+\calcium\ and \proton+\lead. Unfortunately, no experimental data are
available for \proton+\calcium\ at 1~GeV, but since we do not expect a strong
dependence on the projectile energy, we can compare to Chen \etal's data at
763~MeV \cite{chen-ca40}.

Several observations are due. First, the introduction of the neutron skin in
\lead\ boosts the neutron-removal cross section, as expected. This is however
undesired, since the cross section calculated by standard \incl\ is already in
moderate excess of the experimental value. Second, surface fuzziness suppresses
the cross sections for both one-nucleon-removal channels. This is true both for
\calcium\ and \lead. Third, neither effect is sufficient to compensate for the
overestimation of the proton-removal cross section in \lead\
\emph{if considered alone}.

When the two refinements are simultaneously applied to \lead, the effect of
surface fuzziness for neutron removal almost exactly compensates the effect of
the neutron skin, and the final result ($83.8$~mb) is very close to the value
calculated with standard \incl\ ($82.1$~mb), which is within two standard
deviations (about $30\%$) of the experimental value. The proton-removal cross
section, on the other hand, is reduced by almost a factor of two, which brings it
much closer to the experimental datum, but not quite in agreement with it. The
agreement for the \proton+\calcium\ cross sections is also improved: the change
in the proton-removal cross section is minor ($\sim10\%$) and stays within the
experimental error bar, but the neutron-removal cross section is reduced by
$\sim50\%$, in fair agreement with the experimental value.

In conclusion, we have shown that \incl\ fails to describe the cross sections
for one-nucleon removal at high energy. We have used simple shell-model
calculations to show that the key to this deficiency lies in the presence of
neutron skins in heavy, stable nuclei and in the energy content of the nuclear
surface. In the future we will need to generalize our approach to non-magic
nuclei and devise a systematic approach to the description of the properties of
the nuclear surface.

\providecommand{\newblock}{}

\end{document}